\documentclass[aps,prl,twocolumn,groupedaddress]{revtex4-1}
\usepackage{graphicx}
\begin{document}
\title{From latent ferroelectricity to hyperferroelectricity in alkali lead halide perovskites}
\author{Guido Roma}
\email{guido.roma@cea.fr}
\affiliation{DEN-Service de Recherches de M\'etallurgie Physique, CEA,
  Universit\'e Paris-Saclay, F-91191 Gif sur Yvette, France}
\author{Arthur Marronnier}
\affiliation{LPICM, CNRS, \'Ecole Polytechnique, Universit\'e Paris-Saclay, F-91128 Palaiseau, France}
\altaffiliation{DEN-Service de Recherches de M\'etallurgie Physique, CEA, Universit\'e Paris-Saclay,F-91191 Gif sur Yvette, France}
\author{Jacky Even}
\affiliation{Univ Rennes, INSA Rennes, CNRS, Institut FOTON, UMR 6082, F-35000
  Rennes, France}
  
\begin{abstract}
Using first principles calculations we show that several alkali lead halides potentially present collective ferroelectric polarization. This should occur at least at the nanoscale; it could be detected macroscopically provided it is not concealed by lattice vibrations in the temperature range of stability of the cubic perovskite phase. For potassium lead halides and for alkali lead  fluorides, remarkably, the ferroelectric behavior turns hyper-ferroelectric, suggesting a more robust ferroelectric polarization in spite of depolarization potentials induced by charge accumulation on surfaces or interfaces.
\end{abstract}

\maketitle

Ferroelectricity arises from cooperative polar distortions that
stabilise a non-centrosymmetric crystalline structure. 
An electric field beyond a given threshold can collectively flip the
dipole moments associated with the local distortions, which leads to hysteresis
in the polarization/electric field curve. When thermal
fluctuations are sufficient to destroy long range correlations between the
orientation of local dipoles, the
crystal becomes paraelectric, keeping however an enhanced dielectric
susceptibility in a given range of electric fields.
While oxide perovskites have been largely studied for their ferroelectric or
paraelectric behavior\cite{DawberRabeScott_TFoxideFerro_RMP2005,Ghosez2011}, halide perovskites are mostly studied in recent years
for photovoltaic applications, in particular hybrid organic-inorganic ones. In
the latter, the A cation in the ABX$_3$ perovskite structure is a small
organic molecule whose dipole moment, if aligned in a periodic arrangement,
can produce ferroelectricity\cite{Frost_ferroMAPI_NanoLett2014}. 

Early experiments conducted at room temperature showed that the orientations of the molecular cations correspond to a  paraelectric phase for
CH$_3$NH$_3$PbI$_3$\cite{Fan_FerroMAPI_JPCLett2015} (methylammonium lead
iodide, or MAPbI$_3$). Theoretically\cite{Even_disorder_Nanoscale2016},
dipole orientations are shown to be random in the high temperature cubic
phase. But the structural phase transition from the high temperature cubic to
room temperature tetragonal phase of CH$_3$NH$_3$PbI$_3$ is almost impossible to
properly simulate by using {\sl ab initio} molecular dynamics, due to the
limitations of computational ressources. When lowering the temperature, the orthorhombic phase is usually obtained by these methods before the long-range correlations characteristic of the tetragonal phase can be characterized. Recent experiments have shown that MAPbI$_3$ (unlike MAPbBr$_3$) is pyroelectric in its tetragonal phase,  at room temperature and ferroelectricity was further demonstrated at T=204~K  by showing an hysteresis loop in the polarization versus electric field curve\cite{RakitaPNAS2017}. Similar results have been obtained for layered perovskites.\cite{ZhangSum_2Dferro_2019}

Focusing on the
dipole of the organic moiety has somewhat concealed the contribution of the inorganic
framework to the polarization. However, CsPbF$_3$ is an example of an alkali halide perovskite with a
polar, ferroelectric, ground
state~\cite{Berastegui_2001,Smith_CsPbF3phasetrans_InorgChem2015}, and also CsPbI$_3$ presents
polar instabilities in some of its phases~\cite{Marronnier_JPCLett2017,Marronnier_blackphases_ACSnano2018,Sutton2018}, including the cubic, high
temperature, perovskite phase. This phase, or the tetragonal one, are stabilized
at room temperature in the form of quantum dots\cite{Swarnkar2016} or thin
films\cite{WangGraetzel_betaCsPbI3sc18.4_Science2019}, producing highly
efficient solar cells, but the room temperature ground state is, in reality, a
non-perovskite
structure\cite{Marronnier_blackphases_ACSnano2018,Sutton2018}. 

Concerning polar distortions,
 the lack of inversion symmetry in the distorted structure,
together with strong spin-orbit coupling, leads to the Rashba-effect that could
reduce the recombination rate\cite{zheng2015rashba}; exciton binding energy could be very low as a
consequence of large dielectric screening associated to the ionic contribution
to the dielectric constant~\cite{Even_excScreening_JPCC2014,Miyata2015}. The
role of these two propositions on carrier dynamics has been downplayed because, for the former, first
principles calculations show that the influence of the
Rashba effect on recombination is weaker than initially
expected, both in hybrid~\cite{ZhangJPCLett2018,ZhangACSELett2018} and
inorganic~\cite{Marronnier_Rashba_JPCC2019} halide perovskites; for the
latter hypothesis, the calculated ionic contribution to the dielectric
constant\cite{Marronnier_Ellipsometry_SOLMAT2018} is large but not huge.
A third scenario, proposed by Miyata\cite{Miyata2018}, involves polar nanodomains surrounding a ferroelectric large
polaron, which could provide a much more effective screening and a potential
barrier hindering charge recombination. Several works have discussed the
possible role of ferroelectric
distortions\cite{Frost_molecularFerro_APLMater2014,Chen_FerroHOIP_JMaterChem2015},
 highly anharmonic polar modes leading to dipolar random fluctuations larger than stochastic dipolar molecular rotations\cite{Katan_EntropyHPero_NatMater2018} competing with
octahedral rotations\cite{Smith_CsPbF3phasetrans_InorgChem2015,Radha_CsYX3distortions_PRM2018} in halide perovskites, but a detailed knowledge of internal
electric fields and polarization in these materials is still missing and it is necessary in
order to develop further any model relying on a supposed ferroelectricity. 

In this paper we focus on inorganic alkali lead halide perovskites, keeping
Pb in the ABX$_3$ structure because
lead's lone pair seems to be instrumental in stabilizing a polar ground
state\cite{Smith_CsPbF3phasetrans_InorgChem2015}; we show that all
the cubic perovskite phases of APbX$_3$ compounds studied here (A=Cs, Rb, K ;
B=I, Br, Cl, F) present
 a ferroelectric behavior, to various degrees, reaching in some cases the
 hyperferroelectric\cite{Garrity_hyperferro_PRL2014} regime; when
 ferroelectricity, or hyperferroelectricity, is wiped out by thermal
 fluctuations (which is apparently the case for
 the $\alpha$-phase of CsPbI$_3$ in its stability domain) the material is
 paraelectric, or {\sl hyperparaelectric}, but conserves the internal polar distortions
 typical of normal, or hyper, ferroelectrics, now showing up as dynamical
 fluctuations of local dipole moments.

Our approach is based on Density Functional Theory (DFT) calculations of four
crucial quantities along the distortion pattern of a soft polar phonon: the
Kohn-Sham total energy $U$, the
unit cell volume $\Omega$, the
polarization $P$, and the dielectric constant in the high frequency limit
$\varepsilon_\infty$. The dielectric constant is obtained with Density Functional Perturbation Theory\cite{Baroni_DFPT_RMP2001} and the polarization
is obtained with the Berry phase approach\cite{Resta1994}. We used the
Quantum-Espresso package\cite{QE} for the calculation of
the four ingredients of the model\cite{TechDetails}.     

 We describe the atomic displacements along the
soft phonon previously described for CsPbI$_3$\cite{Marronnier_JPCLett2017} with a
parameter, $\lambda$:
we define $\lambda=0$ for the centrosymmetric structure (with space group $Pm\overline{3}m$) and
$\lambda=1$ for the distorted minimum energy configuration. Intermediate
configurations are obtained by linear interpolation between symmetric and
asymmetric structures. We fit $U(\lambda)$ with a
6th order polynomial with even powers of $\lambda$, as in our previous studies
of CsPbI$_3$\cite{Marronnier_JPCLett2017,Marronnier_blackphases_ACSnano2018}.
 For the fitting of polarization $vs$ $\lambda$ we use an odd
polynomial of fifth degree, for the susceptibility an even combination of
Chebyshev polynomials up to 8th degree. 

\begin{figure}
\includegraphics[width=\columnwidth]{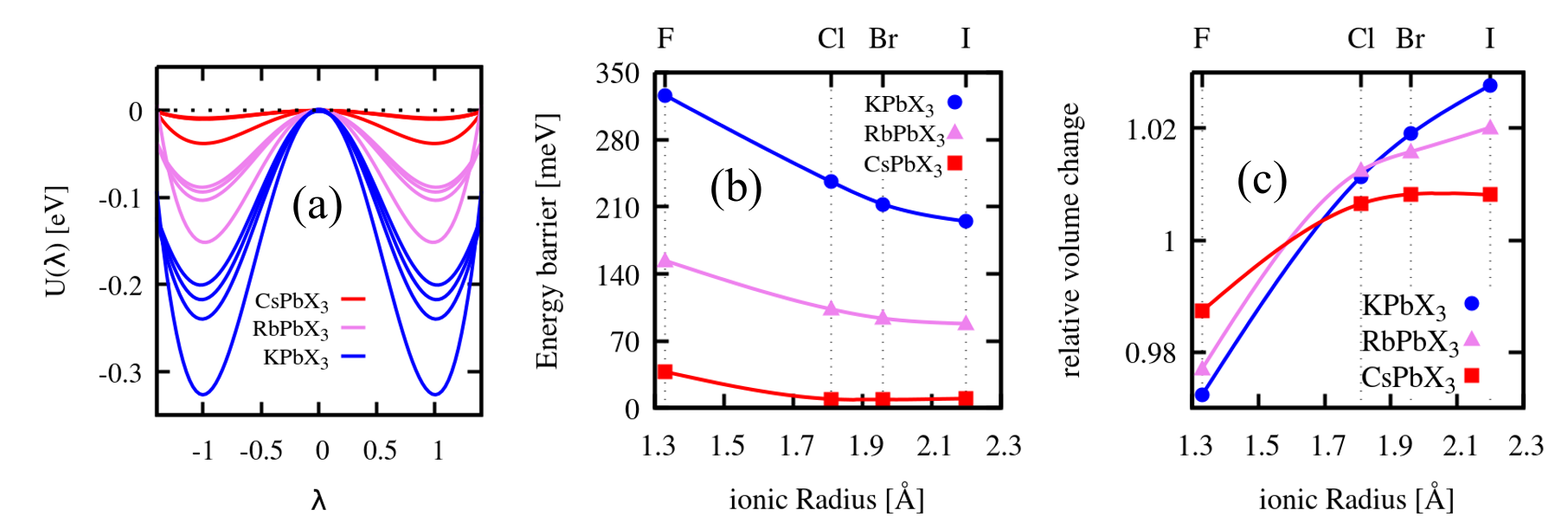}
\caption{ (a) Total energy profiles U($\lambda$), where $\lambda$ parametrizes the ferroelectric distortion along the soft polar phonon
  eigenvector; (b) Potential energy barriers
  U(0)-U(1), of the profiles in (a), plotted $vs$ the
  ionic radius of the halogen atom;\cite{radii} 
  (c) Relative volume change
  between the asymmetric and symmetric configuration: $\Omega(\lambda=1)/\Omega(\lambda=0)$.}
\label{Uprofiles}
\end{figure}

All twelve APbX$_3$ compounds present the same double well profile
of the total energy, U($\lambda$), around the symmetric $Pm\overline{3}m$ structure previously described for
CsPbI$_3$ (see Fig.~\ref{Uprofiles}a). The energy barriers of the double
well range from 9~meV for CsPbBr$_3$ to 326~meV for KPbF$_3$ (Fig.~\ref{Uprofiles}b). The displacement along one of the
three degenerate soft polar phonons (we choose the $x$ direction)  leads to a
tetrahedral distortion and a slight volume expansion for all compounds
except the three fluorides, which contract (Fig.~\ref{Uprofiles}c); this
feature underlines a possible coupling between optical and acoustic phonons.

Coupling between polar and non polar distortions may affect the nature of the
ferroelectric behavior and the presence of a soft polar phonon does not mean
automatically ferroelectricity. A deeper understanding can be gained by
studying the electric equation of state, relating the electric field and
either the
polarization P or the electric dispacement D.
We consider the electric energy functional espression\cite{Stengel2009} for
a solid in an external electric field E, as
proposed in Ref.~\onlinecite{AdhikariFu2019}:
\begin{equation}
F(\lambda,E)=U(\lambda)-\Omega(\lambda)\left[P(\lambda)E+\frac{1}{8\pi}\chi_\infty(\lambda)E^2\right],
\end{equation}
where we assumed atomic units, substituting $\varepsilon_0=\frac{1}{4\pi}$ for the dielectric constant of free space.

In addition to the total energy U and the volume,
$\Omega$, the espression contains also the polarization P and
the electronic susceptibility  $\chi_\infty=\varepsilon_\infty-1$, which we
both calculated for various $\lambda$, $0<\lambda<1.4$,
and fitted them with polynomials.
From these ingredients we can obtain the electric field by solving the second degree equation
minimizing the electric energy functional $\left(\frac{\partial
  F\left(\lambda\right)}{\partial\lambda}=0\right)$\cite{AdhikariFu2019} and,
with it, obtain the parametric P-E and D-E curves, i.e., the electric
equation of state.

\begin{figure}
\includegraphics[width=\columnwidth]{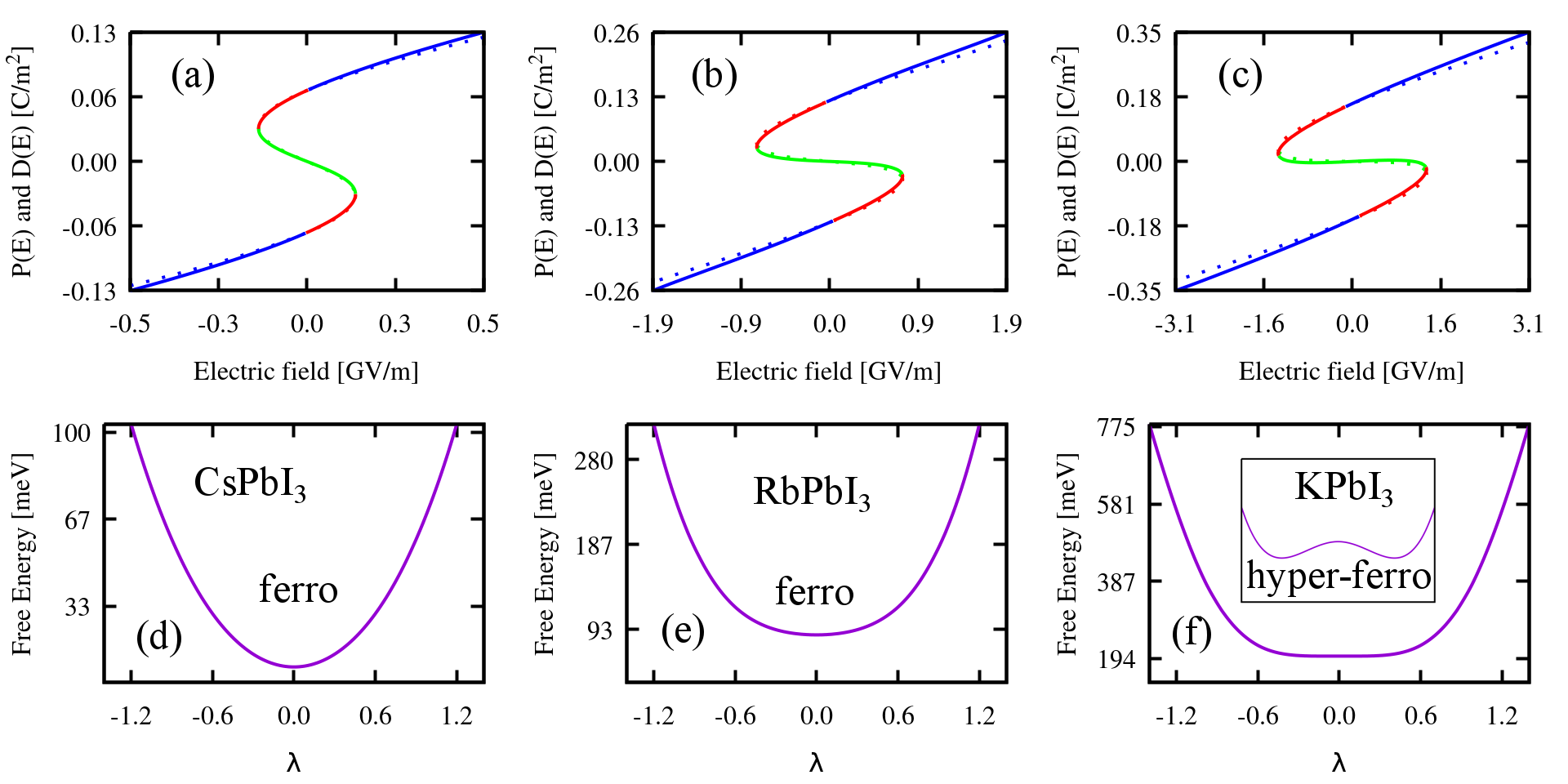}
\caption{Panels (a-c): the electric displacement field (full line) and the
  polarization (dotted line) $vs$ electric field curves for the three alkali
  lead iodides CsPbI$_3$ (a), RbPbI$_3$ (b) and KPbI$_3$. Panels (d-f):  the
  electric energy F in open circuit boundary conditions (OCBC) $vs$ the
  parameter $\lambda$ for the same compounds. While CsPbI$_3$ and
  RbPbI$_3$ behave as a normal proper ferroelectric, with a single minimum in the
  F($\lambda$) curve, potassium lead iodide shows incipient
  hyperferroelectricity. (see the inset in panel (f), where th y-axis spans
  half a meV).}
\label{APbI3}
\end{figure}

Let us consider first the three iodides: 
the parametric P-E (or D-E) curves shown in Fig.~\ref{APbI3}a-c present the typical S-shape associated to the hysteresis in ferroelectric materials,
confirming the potential ferroelectric behavior of these cubic perovskite
phases. We use the red/green/blue colors to distinguish the locally stable,
locally unstable and globally stable regions, respectively, which correspond
to $\lambda$ values where $\frac{\partial^2 U}{\partial \lambda^2}>0$ and
$|\lambda|<1$ (locally stable), $\frac{\partial^2 U}{\partial \lambda^2}<0$
(locally unstable) and    $\frac{\partial^2 U}{\partial \lambda^2}>0$ and
$|\lambda|>1$ (globally stable). The shape of the curves evolves from
CsPbI$_3$ to KPbI$_3$, widening progressively the range of electric fields
spanned by the unstable region.

The Kohn-Sham total energy double well profile, as well as the volume, polarization and
dielectric constant, are obtained with calculations at zero field (E=0) and
within short circuit boundary
conditions (SCDC), which means that the electric field vanishes  at the
hypothetical surface of an infinite bulk described within periodic boundary
conditions. Experimentally, this corresponds to a macroscopic sample where the surfaces
are kept at the same potential. Conversely, in several experimental situations
implementing open circuit
boundary conditions (OCBC), charges accumulate at the surface and induce a
depolarization field which counters the buildup of polarization inside the
sample. In practice, the polar ground state is at least partially destroyed by the
depolarization field. This effect can be predicted by
inspecting the electric energy in
OCBC\cite{AdhikariFu2019}:
\begin{equation}
F(\lambda)=U(\lambda)-\Omega(\lambda)\left\{-\frac{4\pi P(\lambda)^2}{1+\chi_\infty(\lambda)}+\frac{1}{2}\chi_\infty(\lambda)\frac{4\pi P(\lambda)^2}{[1+\chi_\infty(\lambda)]^2}\right\}
\end{equation}
which will show a double well instability only
if the depolarization field is not sufficient to destroy the bulk
polarization.
In panels d-f of Fig.~\ref{APbI3} we show the electric energy in OCBC alongside with the
P(E) and D(E) curves previously discussed.
 The electric energy in OCBC presents a single minimum for CsPbI$_3$ and RbPbI$_3$,
but has still a (very shallow) double well profile for KPbI$_3$. This double well is
the signature of what has been called a
hyperferroelectric~\cite{Garrity_hyperferro_PRL2014}: a material which
is able to keep a bulk ferroelectric polarization in spite of the
depolarization field induced by surface charge accumulation.

\begin{figure}
\includegraphics[width=\columnwidth]{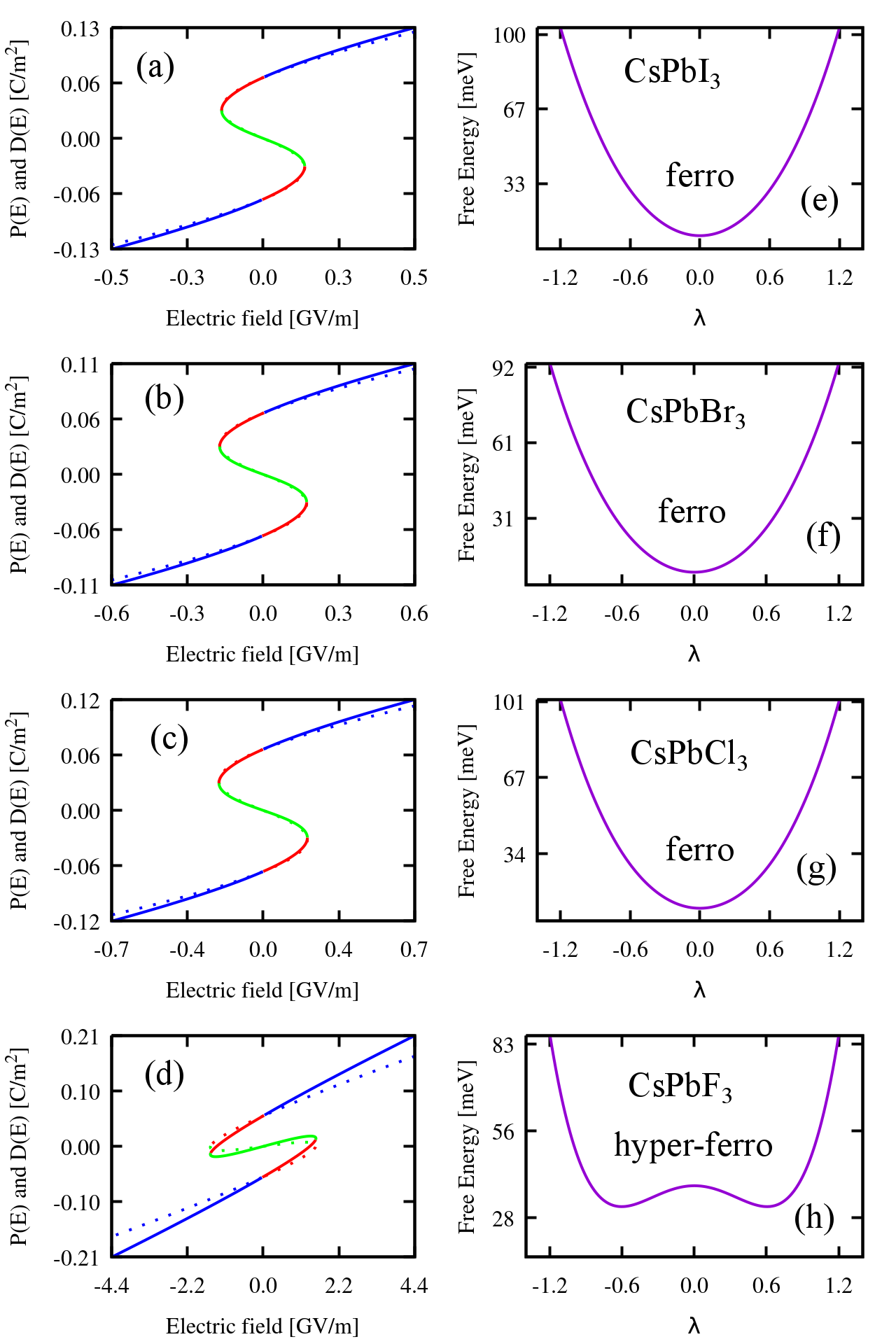}
\caption{The electric displacement field (full line) and the
  polarization (dotted line) $vs$ electric field curves (left column)  and the
  electric energy F in open circuit boundary conditions (OCBC) $vs$ the parameter $\lambda$ (right column) for the four
  caesium lead halides: CsPbI$_3$, CsPbBr$_3$, CsPbCl$_3$ and CsPbF$_3$. While
  the first three behave as normal proper ferroelectrics, with a single minimum in the
  F($\lambda$) curve, caesium lead fluoride presents a clear
  hyperferroelectric character.}
\label{CsPbX3}
\end{figure}

In a similar way we can analyse the trend while varying the halogen atom. For
the sake of illustration we show in Fig.~\ref{CsPbX3} caesium lead
halides, where the evolution from iodide to fluoride is stronger. Starting
from CsPbI$_3$, where the ferroelectric behavior stands on a shallow double
well for U($\lambda$) whose energy barrier is only on the order of 10 meV
(see Fig.~\ref{Uprofiles}a-b), we end up with
CsPbF$_3$, with a barrier almost four times larger and whose already discussed
ferroelectricity\cite{Smith_CsPbF3phasetrans_InorgChem2015} we show
to be, in reality, hyperferroelectricity. In between them, CsPbBr$_3$ and
CsPbCl$_3$ behave very similarly to CsPbI$_3$.

\begin{figure}
\includegraphics[width=\columnwidth]{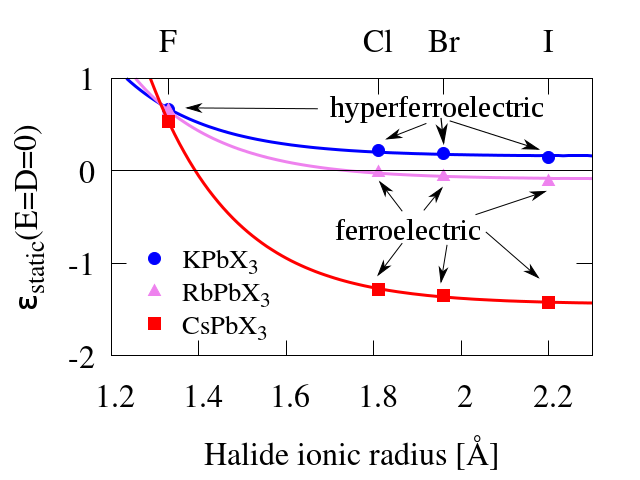}
\caption{Static dielectric constant at zero field, or $\frac{dD}{dE}\left|_{E=0}\right.$ for alkali
  lead halides as obtained from our first principles calculations of the D-E parametric curves, plotted versus the
halogen ionic radius. The lines are a fit of the data with a product of a
quadratic function of the alkaline ionic radius and an exponential function of
the halogen ionic radius.}
\label{staticFerro}

\end{figure}

The hyperferroelectric behavior manifests itself not only in the double well
shape of the electric energy in OCBC, but also in the shape of the D-E
curves (electric displacement $vs$ electric field), qualitatively
very similar to the P-E curves. In normal dielectrics the derivative of the electric displacement $vs$ the
electric field (i.e., the static dielectric constant, $\varepsilon_{static}$) is
positive, including when the external electric field vanishes $\left(\frac{dD}{dE}\left|_{E=0}>0\right.\right)$; at the
ferroelectric transition, however, the dielectric constant diverges and, for a
proper ferroelectric, becomes multivalued (hence comes hysteresis); at the
crossing of the D=0 axis (in the unstable region)
it becomes negative $\left(\frac{dD}{dE}\left|_{E=0}<0\right.\right)$. If we
could progressively tune the shape of the D-E curve from a normal
ferroelectric towards a hyperferroelectric, $\frac{dD}{dE}\left|_{E=0}\right.$
would at some point vanish and become positive in the hyperferroelectric state. 
We consider then $\varepsilon_{static}^{E=0}=\frac{dD}{dE}\left|_{E=0}\right.$
in the unstable region as a measure of the strength of
the ferroelectric polarization, neglecting here the role of lattice vibration,
which will eventually determine whether the material is (hyper-)ferroelectric
or (hyper-)paraelectric.
We use this parameter to follow the trends in the ferroelectricity
of alkali lead halides. In Fig.~\ref{staticFerro} we show the value of $\varepsilon_{static}^{E=0}$
for the twelve alkali lead halides considered in this paper.

These reults reveal, first, that the
trends with alkali substitution, observed in Fig~\ref{APbI3} for the iodides, is
maintained for bromides and clorides. Second, all alkali lead fluorides
considered here are
hyperferroelectric. Third, all potassium lead halides, too, present
hyperferroelectric behavior.

Most of these compounds are found experimentally with the cubic perovskite
structure only at relatively high temperatures:
CsPbI$_3$ $\alpha$-phase is stable only above
373$^\circ$C\cite{Marronnier_blackphases_ACSnano2018,Sutton2018}, CsPbCl$_3$
assumes the cubic perovskite structure
above 47$^\circ$C\cite{Moeller2} and according to dielectric measurements is
not ferroelectric, although Raman measurements show soft phonon
modes\cite{Hirotsu1970}. CsPbBr$_3$ becomes cubic only above
130$^\circ$C\cite{Moeller2,Hirotsu1974}. The fluoride CsPbF$_3$, conversely, is in
the $Pm\overline{3}m$ symmetry at room temperature, transforming to
rhombohedral $R3m$ below -83$^\circ$C. As far as rubidium compounds are concerned, the
bromide has a perovskite structure only above 290$^\circ$C\cite{Tang_RbPbBr3_AngewChem2019} and the fluoride
above 217$^\circ$C\cite{Yamane2008}.
We could not find reports on solid phases of
potassium triiodides, except from
a citation of KPbI$_3$ as occuring in NH$_4$CdCl$_3$-type structure ($Pnma$
symmetry)\cite{Zandbergen1979} 

This tendency to lose the cubic perovskite structure below a certain
temperature can be interpreted in terms of the
so called Goldschmidt tolerance factor, useful to assess the
tendency of an ABX$_3$ compound to stabilize in the cubic perovskite
structure. Assuming Shannon radii\cite{radii}
the three alkali lead iodides have tolerance factors ranging from 0.75 (for
potassium) to 0.85 (for caesium), somewhere at the lower limit to form
perovskite structure. 
A tolerance factor smaller than one indicates that
octahedral tilt rotations can lower the energy of the system, and
this explains why several of the compounds considered in
this paper have an orthorhombic non-perovskite low temperature ground state
and, after intermediate phases, switch to the cubic perovskite structure only
at sufficiently high temperature. A mechanism for the transition, as recently
shown\cite{BechtelVanderVen2019}, is due to the temperature evolution of the
distribution of tilting angles, which from bimodal at low temperature switches
to a single peak at the transitioin to the cubic phase. It has been suggested
that octahedral rotations and ferroelectric distortions are mutually
exclusive\cite{Radha_CsYX3distortions_PRM2018}. The competition between these
two mechanisms depends also on how the
respective double well depths vary with volume expansion. A full understanding of this
point would need a thorough study of
thermal expansion including anharmonic effects in these materials, however a
simpler check is worthwhile: we calculated the depth of the potential well for the
ferroelectric distortion at a few
negative pressures for CsPbI$_3$ and CsPbF$_3$. The results, summarized in Table~\ref{TabPressure}, show
that thermal expansion does not affect the
crucial features of the energy landscape governing ferroelectricity in
these materials.

\begin{table}
\caption{Pressure dependence: $\Delta\Omega$ is the difference between the
  theoretical volume of the
  asymmetric (symmetric) structure and the experimental value at the lowest
  temperature of the cubic phase (554~K for
  CsPbI$_3$\cite{Marronnier_blackphases_ACSnano2018} and 148~K for
  CsPbF$_3$\cite{Berastegui_2001}); $\Delta U$ and $\Delta H$ are energy and
  enthalpy differences between symmetric and distorted configuration. }
\begin{tabular}{l c c c c c }
Halide     & P (kbar) &  $\Delta\Omega$ & $\Delta U$ &
$\Delta H$ & $\varepsilon_{static}^{E=0}$ \\ \hline
CsPbI$_3$  & 0 & -5.7\% (-4.8\%) & 10 & 10 & -1.43\\ 
           & -5 & -3.4\% (-1.8\%) & 8 & 20 & -1.05\\ 
           & -10 & -0.3\% (+2.1\%) & -1 & 36 & -0.88\\ \hline
CsPbF$_3$  & 0 & -2.4\% (-3.7\%) & 38 & 38 & 0.54\\ 
           & -20 & +1.6\% (+2.2\%) & 25 & 33 & 0.57 \\ 
           & -25 & +3.0\% (+4.8\%) & 6 & 36 & 0.95 \\ \hline
\end{tabular}
\label{TabPressure}
\end{table}

Independently from the issue of the stability of the cubic phase for each of
the considered compounds, we remind that doping hybrid perovskite solar cells with
Cs, Rb and, recently,  also K was shown to be beneficial for the stability and
performance of halide perovskite solar cells. In particular, 
potassium doping was shown to improve luminescence\cite{Abdi-Jalebi_KdopedHPERO_Nature2018}, reduce hysteresis\cite{Tang2017},
improve stability and reduce ion
migration~\cite{JHPark_KdopeHPERO_NanoLett2017}, by decorating
surfaces and grain boundaries. Apart from reducing the band gap and inducing potential beneficial elastic effects ---doping with K
tends to shrink the lattice parameter---, we suggest that potassium could enhance charge
separation thanks to a local hyperferroelectric polarization, possibly stabilizing
polar nanodomains responsibles for the
large ferroelectric polaron mechanism\cite{Miyata2018}.

We might expect a ferroelectric behavior similar to alkali lead halides
 in analogous compounds where lead is substituted by other
elements in the same column of the periodic table, like Sn, Ge or
Si\cite{Radha_CsYX3distortions_PRM2018},  suggesting further experimental studies on the
phase stability of these materials.
Finally, we remind that the soft polar phonons which
are at the origin of the ferroelectric distortion may also cause a dynamical
Rashba effect in CsPbI$_3$\cite{Marronnier_Rashba_JPCC2019}, supporting the
possibility of an electric field control of the spin texture in alkali lead
halide thin films.

In conclusion, using first principles calculations we have shown that several
alkali lead halides potentially present collective ferroelectric polarization. This should occur at
least at the nanoscale; it could be detected  macroscopically provided 
it is not concealed by lattice vibrations in the temperature
range of stability of the cubic perovskite phase. For potassium lead halides and
for alkali lead fluorides, remarkably, the ferroelectric behavior turns
hyperferroelectric, suggesting a more robust
ferroelectric polarization in spite of depolarization potentials induced by charge
accumulation on surfaces or interfaces.

\section{Acknowledgements}

This work was granted access to the HPC resources of TGCC under the allocation 2018A0010906018 made by GENCI and under the allocation by CEA-DEN.

\bibliography{FerroPero}

\begin{thebibliography}{44}
\bibitem [{rad()}]{radii}%
  \BibitemOpen
  \href@noop {} {}\bibinfo {note} {We use Shannon ionic radii, assuming
  oxidation state +1,+2 and -1 for alkali, lead and halogens respectively. The
  coordination number is 6 for lead and the halogens and 12 for alkali
  atoms.}\BibitemShut {Stop}%
\end{thebibliography}

\end{document}